\documentstyle[twoside,11pt,aaspp4]{article}
\newcounter{currentfig}

\tighten

\slugcomment{Submitted to Ap.J.}

\lefthead{Thakar et al.}
\righthead{Counterrotating Disk Formation in NGC~4138}

\begin{document}

\title{NGC~4138 - A Case Study in Counterrotating Disk Formation}
\author{Aniruddha R. Thakar, Barbara S. Ryden\altaffilmark{1}}
\affil{Department of Astronomy, The Ohio State University, Columbus, OH
43210-1106\\email: thakar.4@osu.edu; ryden@payne.mps.ohio-state.edu}
\author{Katherine P. Jore, Adrick H. Broeils\altaffilmark{2}} 
\affil{Center for Radiophysics and Space Research, and National Astronomy and
Ionosphere Center\altaffilmark{3}, Cornell University, Ithaca, NY
14853\\email: jore@astrosun.tn.cornell.edu, broeils@astro.su.se}
\altaffiltext{1}{National Science Foundation Young Investigator.} 
\altaffiltext{2}{Present address: Stockholm Observatory, S-13336
Saltsj$\ddot{\rm o}$baden, Sweden}
\altaffiltext{3}{The National Astronomy and Ionosphere Center is operated by
Cornell University under a cooperative agreement with the NSF.}

\begin{abstract}
The Sa(r) galaxy NGC~4138 has been recently found to contain an extensive
counterrotating disk which appears to be still forming.  Up to a third of the
stars in the disk system may be on retrograde orbits.  A counterrotating ring
of H~II regions, along with extended counterrotating H~I gas, suggests that
the retrograde material has been recently acquired in the gas phase and is
still trickling in.  Using numerical simulations, we have attempted to model
the process by which the counterrotating mass has been accreted by this
galaxy.  We investigate two possibilities: continuous retrograde infall of
gas, and a retrograde merger with a gas-rich dwarf galaxy.  Both processes are
successful in producing a counterrotating disk of the observed mass and
dimensions without heating up the primary significantly.  Contrary to our
experience with a fiducial cold, thin primary disk, the gas-rich merger works
well for the massive, compact primary disk of NGC~4138 even though the mass of
the dwarf galaxy is a significant fraction of the mass of the primary disk.
Although we have restricted ourselves mainly to coplanar infall and mergers,
we report on one inclined infall simulation as well.  We also explore the
possibility that the H$\alpha$ ring seen in the inner half of the disk is a
consequence of counterrotating gas clouds colliding with corotating gas
already present in the disk and forming stars in the process.
\end{abstract}

\section{Introduction}
NGC~4138, an early-type spiral galaxy within the Virgo supercluster, has no
distinguishing feature other than a nuclear ring of star-forming H~II regions
similar to rings seen in many other early-type spirals (see, for example,
\cite{pe93}).  Nevertheless, it has recently joined the growing list of spiral
galaxies that are now known to possess {\em two} disks which rotate opposite
to each other.  The spiral/S0 galaxies currently known to have such gaseous
and/or stellar counterrotating disks are NGC~4550 (\cite{rgk92}), NGC~4826
(\cite{bwk92}), NGC~7217 (\cite{mk94}), NGC~7332 (\cite{fif94}), NGC~4546
(\cite{ga87}), NGC~3626 (\cite{cb95}), NGC~3593 (\cite{bc96}) and most
recently, NGC~4138 (\cite{jbh96}, hereafter JBH).  This latest instance of a
counterrotating disk was discovered as part of a survey of ``boring'' spiral
(Sa) galaxies which had no unusual morphological features.  The
counterrotating disk consists of both stars and gas which add up to about a
third of the mass of the primary disk, and it appears to be still forming or
has formed recently, as evinced by the extended counterrotating neutral gas
surrounding the primary disk and the nuclear ring of star-forming H~II regions
which is also counterrotating (assuming that the counterrotating material does
not remain in the gas phase for long).  An H$\alpha$ emission-line imaging
survey of S0 galaxies (\cite{pe87}) previously detected the inner ring of H~II
regions in NGC~4138, which is classified as S0$^+$ by the {\it Third Reference
Catalogue of Bright Galaxies} (\cite{dv91}) although the {\it Revised
Shapley-Ames Catalogue} (\cite{st81}) classifies it as Sa(r).  A follow-up
study (\cite{pe93}) confirmed the existence of the ring, resolving many
``blobs'' into discrete H~II regions.

The examples of counterrotating disks currently known run the gamut from
counterstreaming stars (e.g., NGC~4550) to counterstreaming gas (e.g.,
NGC~7332), and this variety among the specimens, along with the observed
differences in the velocity dispersions of the two components, strongly
implies an external (merger or slow accretion) origin for the retrograde
material.  The presence of ionized gas along with stellar counterrotation
suggests that at least some of this material was in the gas phase when
accreted and formed stars later.  However, in the cases where the retrograde
disk mass is comparable to the primary prograde disk mass, the merger or
accretion theory invoked must explain how the primary disk can remain
unperturbed in spite of the influx of a large amount of stars and gas.
Simulations of major mergers show clearly that the acquisition of a large
number of stars is quite a disruptive process which leaves a remnant that
resembles an elliptical galaxy more than a disk galaxy (e.g. \cite{ba92};
\cite{bh92a}).  This is true even for gas-rich mergers between disk galaxies
that yield a counterrotating disk (\cite{hb91}).  We can therefore rule out
major mergers in which a large amount of dissipationless matter is dumped into
the galaxy within a relatively short time.  This leaves minor gas-rich mergers
and secular gas infall as the processes most likely to produce massive
counterrotating disks without heating up the primary disk inordinately.

Whereas examples of counterrotating gas in S0s and spirals have been known for
some time (see \cite{sg94} and references therein), the unexpected discoveries
of a few massive, stellar counterrotating disks in galaxies previously thought
to be ``normal'' naturally raise the troubling question: how many more have
been missed?  Systematic searches for counterrotation in spiral galaxies are
just getting under way, and so far the results are mixed.  A recent survey of
28 S0's with high signal-to-noise spectra (\cite{kfm96}) revealed no new cases
of stellar counterrotation but uncovered counterrotating gas in almost a
quarter of the sample.  This implies that most if not all counterrotating
disks must start out as gas disks, and suggests that star formation may
somehow be inhibited in the retrograde gas.  Furthermore, most of the
counterrotating gas disks are substantially smaller than the corresponding
primary disks.  These findings argue against a high frequency of massive,
stellar counterrotating disks in S0 galaxies.  Theoretical studies of the
origin of massive counterrotating disks are necessary to complement the
observations and make some predictions which can be checked against future
observations.

We have been investigating the origin of such massive counterrotating disks in
spiral galaxies with numerical simulations in which we try to produce a
massive counterrotating disk in a fiducial spiral galaxy (\cite{tr96},
hereafter TR).  However, given the large number of input parameters that
can have an appreciable effect on the outcome, it is of great practical value
to adopt a phenomenological approach and study individual galaxies so that we
can narrow down the range of parameters that are relevant to existing
counterrotating disks.

As in TR, we investigate two mechanisms here for producing a counterrotating
disk in NGC~4138.  The first process is gradual or secular gas infall, and the
second is a gas-rich dwarf satellite merger.  Minor mergers and accretion have
been receiving increased attention in the last few years, due partly to the
fact that major mergers have been shown, through numerical simulations, to be
unable to produce some of the observed features in spiral galaxies, and
certainly can be ruled out in cases where cold, thin disks still dominate the
galaxy morphology.  The picture that is emerging is that the importance of
accretion and minor mergers as a means for spiral galaxy evolution may have
been underestimated.  The secular evolution of galaxies most likely is driven
by such events from later types towards S0 (e.g. \cite{wmh96}).  There is also
some evidence from numerical experiments to support the idea that spiral disks
are more resistant to minor mergers if the angular momentum of the merger
orbit is high enough for the primary disk to respond in a coherent manner by
tilting rather than warping or heating (\cite{hc96}).

The prevalence of counterrotating gas disks is clearly indicative of the
importance of gas in the process(es) responsible for forming them.  Indeed, TR
found it quite impossible to produce a counterrotating stellar disk within a
Hubble time from a dissipationless merger without wrecking the primary disk.
However, it is not clear whether it is necessary for all the retrograde matter
accreted to be in the form of gas.  A gas-rich dwarf merger may work nearly as
well as pure gas infall in some cases, especially if the dwarf has a low M/L
ratio.  There are some difficulties associated with secular gas infall,
particularly the source of the infalling gas and the need for the gas to
maintain retrograde angular momentum with respect to the primary disk over a
good fraction of a Hubble time.  Alternatives to gas infall, of which gas-rich
dwarf mergers is the most promising, must be thoroughly investigated to
determine their ability to produce massive counterrotating disks.

Based on numerical experiments with a cold and thin primary, TR found that gas
infall was able to produce massive counterrotating disks without excessive
heating of the primary disk, but the same could not be said of gas-rich dwarf
mergers, which were found to be problematic.  For the significantly hotter and
more compact primary of NGC~4138, gas-rich dwarf mergers do a much better job,
and in fact produce a counterrotating disk in a significantly shorter time
than continuous infall.  We briefly describe our numerical method in section
2.  Our simulation results are presented in section 3.  Analysis of these
results and comparison with previous work in galaxy interactions are in
section 4.  Section 5 lists our conclusions.

\section{Computational Method}
The numerical method used is described in detail in TR.  We use a tree code to
compute the gravitational forces for all particles and a sticky particle
dissipation scheme for the gas dynamics.  We started with a basic version of
the tree gravity solver (\cite{bh86}) written in the C language and optimized
it for execution on the Cray Y-MP at the Ohio Supercomputer Center.  To this
we added a sticky particle code which implements a simple form of gas dynamics
via inelastic collisions which dissipate kinetic energy and conserve linear
and angular momentum.  Our sticky particle code is a combination of previous
schemes (\cite{br77}; \cite{sw81}; \cite{nw83}), and is also optimized for the
Cray Y-MP.

For the simulations described here, we use twice the number of particles to
represent the halo (32k instead of 16k as in TR).  The softening length is
smaller (0.5 kpc instead of 1 kpc used by TR) to give a better resolution.
The primary stellar disk is represented as before with 32k particles.  It is
dissipationless initially and contains no gas (except for infall simulation
I6, see below).  The number of gas particles is also the same, being 20k for
gas infall models and 15k for dwarf merger models.  The dissipationless matter
in the dwarf galaxy is represented by a varying number of particles depending
on its mass.  The mass of each dissipationless particle in the dwarf is
comparable to the mass of each particle in the primary disk ($\sim6\times10^5
M_{\sun}$).  The total number of particles is 84k for the infall simulations
and 86k for the dwarf mergers (except one large $N$ dwarf merger which
contains 162k particles).

Although the two-component model for the spiral galaxy is the same as before,
the model parameters are different since we are modelling NGC~4138 instead of
a fiducial spiral galaxy.  The parameter values for our model of NGC~4138,
based on the observations and mass modelling described by JBH, are summarized
in Table~\ref{dhpars}.  The disk in this case is radially much hotter than the
disk used by TR, and the value of $Q_{\rm o}$ (the value of the stability
parameter $Q$ at $R_{\circ}=8.5$ kpc) we use is 4.0.  This was necessary for
the model rotation curve and primary disk velocity dispersion values to be
consistent with the ``observed'' rotation curve and velocity dispersions of
NGC~4138 obtained by JBH (Fig.~\ref{rcinit}).  Although the initial galaxy
prior to the accretion of the counterrotating material was in all probability
different from the one we see today, we had no choice but to start with a
model as closely resembling the observed post-accretion galaxy as possible.
To investigate what happens if we start with a radially cold primary instead,
we ran one dwarf merger simulation with a $Q_{\rm o}=1.5$ primary disk.

The models for continuous infall and the gas-rich dwarf merger are basically
the same as those used by TR.  For gas infall, a long (length = 150 kpc, width
= 10 kpc), rectangular slab of gas is allowed to fall into the primary spiral
galaxy on a retrograde orbit.  We chose this model because it allows us to
impart the appropriate angular momentum to the gas in a simple manner, and
because we do not really know what the characteristics of real gas infall are.
Our long column of gas (which breaks up into clumps before it falls into the
primary galaxy) can be thought of as tidal debris left over from a previous
interaction.

The initial conditions and some physical parameters of the infalling gas are
different in the present simulations from those used by TR.  The velocity
dispersion of the gas is higher than before ($\sim$30 km/s instead of $\sim$10
km/s used in TR), the mass and initial angular momentum of the infalling gas
are different to suit the current problem, and we have experimented with the
``stickiness'' of the gas to assess its impact on the final outcome.  We have
also studied the effect of an inclined infall orbit and the presence of a
small amount of preexisting prograde gas in the disk.  The parameter values
for the different infall simulations (dubbed I1-I6) are listed in
Table~\ref{infpars}.  The magnitude of the initial angular momentum, $J_z$, of
the gas can be altered by changing the initial velocity of the gas $V_0$ or
the initial distance $R_0$ of the gas slab from the center of the primary
disk, or both.  The total mass of the gas falling in is taken to be 50\% of
the mass of the primary disk, even though the counterrotating mass observed is
at most 30\% of the mass of the disk.  The extra gas is included to allow for
some gas which is dispersed over large distances and does not fall into the
primary disk by the end of the simulation.

Similarly, for the dwarf mergers, the total amount of luminous mass
(gas+stars) in the dwarf galaxy is chosen to be 50\% of the primary disk's
mass.  The dwarf galaxies we have chosen have higher gas contents than do
typical dwarfs.  Our values for $M_g/M_T$, the ratio of the gas mass to the
total mass, range between 0.27 and 0.53, whereas the typical range for
$M_{HI}/M_T$, the ratio of neutral gas mass to total mass, is $0.09-0.24$ for
dwarf galaxies (\cite{rh94}).  However, the latter does not take into account
the molecular gas content in dwarf galaxies, which is not very well quantified
so far.  The mass model for the dwarf galaxy is a spherical Plummer model with
the density profile being a polytrope of index 5 (as described in TR).  The
parameters for the dwarf galaxy in each of the dwarf merger simulations
(M1-M3) are listed in Table~\ref{dmpars}.  The total mass of the dwarf galaxy
therefore ranges between $\sim9$\% and $\sim18$\% of the mass of the primary
galaxy.  Although the physical parameters for the dwarf galaxy are identical
for M2 and M3, the latter uses a radially colder primary disk and a larger
number of particles in the simulation (64k each for the disk and halo, and 34k
for the dwarf galaxy) to minimize numerical heating even further.

In all the simulations presented here, the primary disk is held frozen for
$t=0-0.2$ Gyr, which amounts to $\sim5$ dynamical times.  This is to allow the
halo to virialize.  The disk and halo are then evolved in isolation for
$t=0.2-0.5$ Gyr so that they may achieve equilibrium.  The gas column (for
infall) or the dwarf galaxy is introduced at $t=0.5$ Gyr.  In most cases, the
simulation results are shown starting at $t=0.9$ Gyr, so that the simulation
is already in progress.  The initial time-step is not shown.

With the simple gas dynamics scheme we are using, we do not attempt to study
the internal structure of the gas disks formed in our simulations, and hence
we follow each simulation only up to the point where a flat disk-like
structure with retrograde orbits is formed.  The sticky particle scheme does
not give an accurate description of the detailed gas dynamics, so we restrict
ourselves to studying the overall characteristics of the objects such as the
size and shape of the counterrotating disk formed (if any) and the impact of
the accretion or merger on the primary disk.  For the same reason, we also do
not study the effect of including large quantities of preexisting gas in the
primary disk with this method.  We hope to report on these aspects in the near
future using SPH simulations which will provide a more realistic description
of the gas dynamics with the inclusion of shocks, viscosity, and eventually
even star formation.

\section{Results}
\subsection{Continuous Infall}
Figs.~\ref{ci1}-\ref{ci6} depict the behavior of the infalling gas for the
different infall simulations (I1-I6).  I1 and I2 compare large initial angular
momentum ($J_z$) infall with small initial angular momentum infall
respectively for a moderately dissipative gas (coefficient of restitution
$\alpha = -0.5$).  I3 shows the effect of very low initial gas angular
momentum on the production of a counterrotating disk.  I4 illustrates infall
along an orbit which is considerably tilted with respect to the plane of the
primary disk.  I5 is the same as I1 but with a different parameterization of
the sticky gas which produces a smaller counterrotating disk.  Finally, in I6
we show the effect of placing a prograde ring of gas in the disk in addition
to the infalling gas.  The mass and initial density of the infalling gas are
the same in all simulations reported here.

For I1, the large $J_z$ case (Fig.~\ref{ci1}), the resulting counterrotating
disk is more than twice as large as the primary disk.  The effect of lowering
the initial angular momentum of the gas can be seen in I2 (Fig.~\ref{ci2}).
This approaches the minimum retrograde $J_z$ that the gas can have and still
yield a counterrotating disk.  Although the counterrotating disk formed is
slightly smaller in I2 than in I1, it is still significantly larger than the
primary disk.

If $J_z$ for the gas is reduced further, a counterrotating disk does not form.
(Fig.~\ref{ci3}).  The evolution of the gas is now very different from the
previous two simulations.  A shell forms at $t=1.7$ Gyr, and thereafter the
gas falls rapidly back into the primary disk.  The infall proceeds at a
significantly faster rate, and the side view shows that the end result at
$t=3.7$ Gyr resembles a spheroid rather than a disk (only the $x-z$ plane view
is shown, but the $y-z$ plane view is almost identical).  Fig.~\ref{vfcmp}
provides a comparison of the velocity fields of the primary disk and the gas
for the I2 and I3.  The velocity field of the gas is not ordered for I3, and
it is hard to discern a well-defined orbital direction.  The difference also
shows up in the comparison of the specific angular momenta of the primary and
secondary disks for the two simulations (Fig.~\ref{amcmp}).  Even though the
net angular is retrograde for I3, we are clearly getting out of the regime
which yields a counterrotating disk.

To investigate the impact of a non-coplanar or inclined infall orbit, we
performed one simulation in which the infall proceeds along an orbital
inclination of $\sim30^{\circ}$ with respect to the plane of the primary disk
($x-y$ plane).  The results are shown in Fig.~\ref{ci4}.  A few shells can be
seen forming early on in the gas, and the later panels show that although a
counterrotating disk does appear to form, it is not aligned with the primary
disk even by $t=5.7$ Gyr.  This is evident from the side views, in which the
gas counterrotating disk and the primary disk appear to be inclined by almost
$45^{\circ}$ with respect to each other.  Furthermore, the primary disk flips
by almost $90^{\circ}$ not once but {\em twice}, each flip taking $\sim1.5$
Gyr to accomplish.  It is clear that a near-coplanar orbit is necessary for
the counterrotating gas to settle quickly into the plane of the primary disk.
TR experimented with a small orbital inclination ($5^{\circ}$) and found it to
have no appreciable impact on the outcome.

Although simulations I1-I4 illustrate the general aspects of infall
simulations adequately, they use a sticky particle formulation in which the
dissipative characteristics of the gas are less than ideal.  The resulting
counterrotating disk is considerably larger than the primary disk regardless
of the initial angular momentum.  The radial distribution of the particles
also shows a clear underdensity of particles at the inner radii
(Fig.~\ref{rhgcmp}).  Since we observed this behavior even when the gas was
made very ``sticky'' (by choosing a smaller coefficient of restitution
$\alpha$), we previously concluded that this was a limitation of our infall
model (\cite{th96}) rather than an artifact of the sticky particle model
chosen for the gas.  However, upon choosing a smaller maximum radius for
selecting collisional neighbors (instead of a smaller $\alpha$), the resulting
counterrotating disks were appreciably smaller than before.  This parameter
change has effectively made the gas more dissipative by strongly favoring
collisions between nearby particles (which on the average have higher relative
velocities due to their mutual gravity) over those between remote particles,
even though the total number of collisions is reduced because of the smaller
number of nearest neighbors found for each gas particle.  Since preferentially
selecting collisions between nearby particles is physically more realistic, we
adopt this parametrization for all the remaining simulations discussed below.

The results for the new sticky particle formulation are shown in
Fig.~\ref{ci5} for I5, for which all the other inputs are the same as I1.  A
comparison of the primary disk shown in the first panel of the side view in
Fig.~\ref{ci5} with the counterrotating disk at $t=4.1$ Gyr indicates that the
two are approximately the same size, with the counterrotating disk being
slightly smaller.  The enlarged views for the last time-step reveal some of
the structure of the disk.  The gas is distributed fairly evenly with radius,
and most of it is within the radius of the primary disk.  The counterrotating
disk clearly does not have an exponential profile, and it is tilted by nearly
$45^{\circ}$ to the original plane of the primary disk.  The primary disk also
tilts by a comparable amount, as can be seen in the side views of the primary
disk.  It is not clear what causes the primary disk to tilt so much,
especially when it does not tilt nearly as much when prograde gas is included
in it (see below).

The nuclear ring of H~II regions is a prominent feature of NGC~4138, but the
discovery that it is also counterrotating was a surprise.  Although nuclear
star formation rings in early type spirals are not unusual, this is the first
counterrotating star forming ring known.  Could this ring be a result of
collisions between corotating and counterrotating gas clouds?  If so, why do
the resulting star forming regions end up counterrotating?  We placed a
prograde ring of gas, with one-tenth the mass of the infalling gas, at the
same location as the observed ring and ran the infall simulation I6 with
everything else being the same as for I5 above.  The results are shown in
Fig.~\ref{ci6}.  The apparent thickness of the counterrotating disk in the
side view is due to its projected inclination.  The only discernible
differences between the gas disk formed in this case and the one formed in the
previous simulation (without prograde gas) are the size and the inclination of
the counterrotating disk with respect to the original plane of the primary.
Without prograde gas, the disk is a bit larger and tilts by a considerable
amount, whereas with prograde gas, the disk is smaller and suffers much less
tilting.  It is reasonable for the prograde gas to rob the counterrotating gas
of kinetic energy and thereby produce a smaller counterrotating disk, but what
could the prograde gas have to do with the tilting?  We are not quite sure,
but some arguments on this subject are presented in \S{4.1}.

\subsection{Gas-Rich Dwarf Merger}
The dwarf merger model we have used is described in detail in TR.  We present
the results of three gas-rich dwarf mergers here, the first with very little
dark matter (M1) and the remaining two with a moderate amount of dark matter
(M2 and M3).  All mergers are coplanar and the dwarf galaxy models are
identical except for the amount of dark matter.  M3 uses the same dwarf galaxy
model as M2, but has a colder primary disk and twice the number of
collisionless particles as M2..

In the first dwarf merger (M1), shown in Fig.~\ref{dm1}, we see a
counterrotating disk forming by $t\sim3.1$ Gyr.  Although particles still
appear to be falling in at $t=4.5$ Gyr, these are mostly stellar particles
from the dwarf, and almost all of the gas particles have fallen into
retrograde orbits around the primary disk by this time.  The formation of a
tidal bridge and two tidal tails is evident from the top view in
Fig.~\ref{dm1}.  The primary disk tilts in response to the merger, and the
counterrotating disk which forms is consequently also tilted with respect to
the original plane of the encounter.  The sizes of the primary and
counterrotating disks are comparable, although there are a few gas particles
at larger radii.

With the additional dark matter in the second dwarf merger (M2), the
counterrotating disk is produced much faster (Fig.~\ref{dm2}).  There is less
tidal stripping of the dwarf, although as in M1, two distinct tidal tails are
formed.  By $t\sim2.5$ Gyr, most of the gas is within the radius of the
primary disk, but it has not quite made it to the plane of the disk yet.  The
gas disk becomes steadily smaller after this, and a comparison of the panels
for the last time-step in M1 and M2 shows that the gas disk in M2 is more
compact and less inclined to the original plane of the primary.  However, even
by $t=4.5$ Gyr, the counterrotating gas does not appear to be very flat, and
its apparent thickness is not due to a projected inclination.

The two dwarf mergers are compared in Fig.~\ref{dmcmp}.  Although the
counterrotating disk is tilted in M1, it is not as thick as it appears because
it also tilted in the $y-z$ plane.  Such is not the case for the
counterrotating disk in M2, which actually is thick.  The second merger
produces a puffier counterrotating disk whose velocity field is not completely
circularized.  A comparison of the angular momenta of the two disks shows this
more clearly.  The angular momentum for the counterrotating disk in M2 is
significantly lower than that for M1.  The primary disk in M2 also gains a
pronounced bar (not shown), but does not appear to tilt nearly as much as in
M1.

The stellar component of the dwarf galaxy also attempts to participate in the
counterrotating disk formation, but with much less success.  The stellar
particles are much more spatially dispersed in both M1 and M2, although there
is some flattening discernible in both simulations.  In M1, the stellar
particles also follow the inclination of the primary disk.  The velocity
fields of the stars from the dwarf, however, are not at all regular or
circularized.  In general the angular momenta of the dwarf stars are much
lower than those of the counterrotating gas particles, but they are definitely
of the same polarity as the gas.  So some of the stars do in fact form a
counterrotating stellar disk, with more success in M1 than in M2.

The faster pace of M2 apparently leaves less time for the primary disk to tilt
in response to the incoming material.  The presence of the extra dark matter
fosters a significantly thicker counterrotating disk than in M1.  Surprisingly
(see \S{4.2}), the thickness of the primary disk is not affected much by the
additional dark matter in M2.

For the third merger simulation (M3), the merger details are the same as for
M2, except that the dwarf galaxy is introduced at $t=1.0$ Gyr instead of
$t=0.5$ Gyr.  This is because the colder primary in M3 requires more time to
achieve equilibrium since it is more susceptible to local axisymmetric
clumping.  The final characteristics of the counterrotating disk are the same
as in M2.  The main difference between M3 and M2 is in the heating of the
primary disk, and this is discussed in the following section.

\section{Discussion}
\subsection{Size, Shape and Orientation of Counterrotating Disks}
Both infall and mergers produce counterrotating disks with sizes which compare
well with the observed size of the counterrotating disk in NGC~4138, in terms
of mass as well as extent.  In some cases, the counterrotating disk is
slightly larger in extent than the primary disk, but this is not a big issue
because we usually only follow our simulations up to the point where most of
the gas falls into the primary disk.  The size of the gas disk reduces slowly
with time, but we do not wait until it matches that of the primary very
closely (to conserve our limited CPU resources, and also because we do not
have star formation in our simulations).  The mass of the counterrotating disk
is usually $\sim30$\% of the mass of the primary disk, since we start with a
total secondary mass (in gas for infall and gas+stars for mergers) of
$\sim50$\% of the primary disk mass and $\sim40$\% of it is lost in material
dispersed over a large volume around the primary galaxy (see below).  This is
close to the upper limit of the mass of the counterrotating disk derived by
JBH.

In the infall simulations, most of the dispersed gas accounts for a very small
fraction ($\gtrsim5-10$\%) of the total gas mass.  However, we invariably find
one or two large clumps of gas, with masses in the range $\sim10-30$\% of the
total gas mass, flung to large distances (100-200 kpc) from the primary disk.
The reason for this is that the initial column of infalling gas breaks up into
clumps, the last one or two of which are massive enough to survive the tidal
field of the primary disk.  This also explains why it does not happen in the
dwarf merger simulations, although there is a small amount of gas which is
spread out over a large volume even in these.  The dispersed stars from the
dwarf account for the remaining mass which does not form part of the
counterrotating disk by the end of the simulations.

We do not see a consistent radial profile for the counterrotating disks formed
in our simulations.  The radial profiles shown in Fig.~\ref{rhgcmp} are not
exponential.  The profile for I6 shows a sharp peak at the location of the
prograde gas, which has been included.  It also shows a higher density for gas
at radii above and below the range where the prograde gas is ($\sim1.5\pm0.5$
kpc).  In all the counterrotating disks formed, we see a drop in the density
in the nuclear region of the disk.  This is true even for I6, so the presence
of a small amount of prograde gas does not help.  The fact that the size of
this ``hole'' is of the order of the gravitational softening length may be a
clue as to its cause.  This may also be an indication of the lowest angular
momentum achievable by the gas as a result of insufficient dissipation in our
models (see below).  It should be noted, however, that the H~I data for
NGC~4138 also show a depression in the center (Fig.~3 in JBH).

The thickness of the counterrotating gas disks is also a matter of concern.
We find that in the majority of our simulations, the counterrotating disk is
somewhat thicker than the primary disk. This is especially so for M2 (and M3).
Moreover, the thickness does not reduce appreciably with time on scales of a
few dynamical times.  One factor that may be causing this is that our gas
starts out with a higher velocity dispersion than in TR, and it does not cool
as much as it should.  There is a cutoff for the collisions for relative
velocities of less than 10 km/s, which is quite low and should allow the gas
to cool down.  However, if most of the cooling takes place before the gas disk
takes shape, the gas will undergo much fewer collisions in the disk state.
Alternatively, the thickness may be due to some other limitation of the gas
dynamics used.

Except for the inclined infall simulation, in all other cases the
counterrotating disk seems to faithfully follow the orientation of the primary
disk, even in cases where the primary disk tilts by a considerable amount.
This is reassuring but not too surprising, since most of the change in the
orientation of the primary seems to occur after the counterrotating disk is
mostly formed.  The largest disk inclination is seen in infall simulation I5.
A calculation of the torques acting on the disk due to the halo and the gas
reveals that the rate of change of the disk's angular momentum matches the
total torque acting on it.  Neither the halo nor the gas can provide enough
torque individually to explain the change in the disk's angular momentum.  The
total torque acting on the disk fluctuates considerably on time-scales as
short as a few Myr.  We have also calculated the axial ratios and axial
vectors of the halo at different times by diagonalizing the normalized inertia
tensor, \begin{displaymath} M_{ij} = \sum{m x_i x_j}, \end{displaymath} where
the $x_i$ are the components of the unit vector parallel to the radius vector
of the particle. There is a slight flattening of the halo along the $z$-axis,
and the axial vectors change with time in a manner consistent with the
inclination experienced by the disk.  The inclination of the disk therefore is
intimately linked to the details of the dynamical evolution of the halo and
the evolving gas, and different initial conditions are apt to result in very
different behavior of the disk.

\subsection{Impact of Accretion on Primary Disk}
If a counterrotating disk forms which resembles the observed disk, then the
next question is whether the primary disk suffers a lot of heating as a result
of the accretion of the new disk via either infall or a merger.  A measure of
the heating of the primary disk is its mean thickness as a function of radius.
We have plotted the mean half-thickness in Fig.~\ref{zthcmp} for the important
infall and merger simulations.  As explained in TR, the primary disk is also
heated in the absence of any accretion due to the discreteness of the particle
representation and two-body encounters with halo particles.  This effect must
be taken into account when trying to measure the thickening due to the infall
or merger.  We have shown a thickness plot for the isolated primary disk
evolved up to $t=4.5$ Gyr in addition to the post-accretion plots in
Fig.~\ref{zthcmp}.  Within $\sim 4$ kpc, where the thickness measurements are
more reliable, the total increase in thickness is not much higher than that
due to the numerical heating.  When the numerical heating is reduced, as in
the case of simulation M3, we see a sharp drop in the disk thickening.  Since
the two effects are not strictly separable, it is hard to make a quantitative
statement about the thickening due to accretion only.  The best we can say is
that the average increase for $R\lesssim4$ kpc is at most of the order of the
numerical thickening.

Whereas the results for infall are consistent with the findings of TR, a
comparison of Fig.~\ref{zthcmp} with Fig.~17 in TR reveals that the gas-rich
dwarf merger performs much better for NGC~4138 than for the fiducial primary
in TR.  This is especially remarkable considering that the dwarf galaxy in TR
was only $\sim4$\% as massive as the primary galaxy, with its luminous mass
being $\sim10$\% of the mass of the primary disk, whereas the corresponding
figures for the dwarf in M1 are $\sim9$\% and $\sim50$\%.  In M2, the
total mass of the dwarf is $\sim18$\% of the mass of the primary galaxy.  Thus
the dwarf galaxies chosen for the NGC~4138 simulations are considerably more
massive relative to the primary galaxy than in TR, and the merger time-scales
are also much shorter.

The secret to the success of the dwarf mergers in NGC~4138 lies in the nature
of its primary disk.  The compact and massive primary disk, due to its higher
surface density, has a higher vertical velocity dispersion and is therefore
initially thicker and much more resistant to vertical heating than the primary
used in the simulations by TR.  The ratio of scale length to scale height for
NGC~4138's disk is only $\sim5$, compared to $\sim10$ for the disk used by TR.
We chose to model a disk that was radially hot as well because it matched the
observed parameters of NGC~4138's current disk well.  Of course, the primary
disk could have been quite different prior to the formation of the
counterrotating disk, and may have been significantly colder radially.  This
case is represented by M3, where we start with a radially colder disk, and use
twice as many collisionless particles to minimize numerical heating.  The
post-merger disk is now even closer to the observed NGC 4138 disk since it
suffers less vertical heating along the way.

\subsection{Comparison with Observations}
The lack of remnants of any tidal features in the vicinity of NGC~4138 may be
used as an argument against a dwarf merger origin for the counterrotating
disk, but it is not a serious problem.  If the merger takes only a few Gyrs to
complete, as we have shown, then there has been plenty of time for any tidal
features to be absorbed by the primary by now.  The extended neutral gas may
represent the last of the tidal tails falling back into the primary.  The
higher mass of the dwarf speeds up the merger considerably.  Low-mass
satellites would not have been able to spiral in within a Hubble time,
especially on a retrograde orbit (\cite{bt87} and \cite{wmh96}, Fig. 13), nor
would a single low-mass merger have been able to provide enough mass for the
counterrotating disk.

There is nothing in the observations of NGC~4138 to date that precludes gas
infall as a means of obtaining a counterrotating disk.  The extended neutral
gas observed could be the ``smoking gun'' from the infall process which
produced the counterrotating disk.  Although there is some uncertainty about
the mass of this extended gas, it is not likely to be more than a few percent
of the mass of the counterrotating disk, which is consistent with our infall
simulations I5 and I6, where the mass of gas around the primary galaxy amounts
to a few percent of the mass of the counterrotating disk.  The large gas
clumps we see at further distances in both of these simulations are more
problematic, since no counterpart is observed, at least in H~I gas, around
NGC~4138 (JBH).  They owe their existence at least in part to our initial
conditions, which allow the tail end of the infalling gas to form clumps dense
enough to resist the tidal influence of the primary.  Reducing the initial
length of the infalling slab of gas by leaving out some of the gas would
produce smaller or no leftover clumps of gas.

The presence of A stars along with the absence of O and B stars from the
primary disk (JBH) suggests that star formation took place less than a Gyr ago
in the primary but ceased thereafter.  Does the influx of counterrotating gas
suppress star formation in the primary?  Our simulations indicate that large
quantities of counterrotating gas make their way inside the primary disk only
at later times, within the last $1-1.5$ Gyr of the simulation.  However,
keeping in mind that this may be due to a limitation of our models, we cannot
rule out the presence of counterrotating gas in the primary disk when the last
burst of star formation took place.  Although galaxies which harbor
counterstreaming gas do not appear to have any ongoing star formation (e.g.,
\cite{fif94}), there is no direct proof to date that counterstreaming gas
inhibits star formation.

On the other hand, since a small amount of primordial prograde gas in the
primary does not cause an appreciable inflow of gas to the center (simulation
I6), we are more inclined to believe that collisions between counterstreaming
gas clouds trigger star formation rather than inhibiting it.  Incoming
counterrotating gas would generally have the larger specific angular momentum,
as demonstrated in Fig.~\ref{ampro}.  The direction of the orbits acquired by
stars resulting from the interaction between corotating and counterrotating
gas would depend upon the relative masses of the two gas components.  A more
massive prograde component should produce significantly more stars with
prograde orbits.  Small amounts of prograde gas would produce stars with
mostly retrograde orbits.  In either case, some gas will flow inwards.

Under the above hypothesis, the A stars seen in the primary disk are an
indication that counterrotating gas was present throughout the primary disk
less than a Gyr ago.  The star formation would have proceeded inwards as more
and more counterrotating gas found its way there, ending earlier at radii
where prograde gas was exhausted faster.  The annulus of counterrotating G and
K giants around the H$\alpha$ ring supports these ideas.  The radial
distribution of corotating G and K giants reflects the primordial radial
distribution of prograde gas (decreasing smoothly from the center outwards).

If the counterrotating gas owes its origin to a merger instead of infall, at
least some of the counterrotating G and K giants could have been inherited
from the dwarf galaxy.  In our dwarf merger simulations, we found that even
though the stars from the dwarf have net counterrotation, they form an oblate
structure which resembles an extended (radius $\gtrsim3$ times radius of
primary) and rather thick (scale height $\sim2-3$ kpc) disk, rather than the
compact and flat disk formed by the gas.  This distinction is clearer in M1
where the gas disk is flatter.  Observationally, such an extended
counterrotating stellar population may be hard to separate from the halo stars
and hence missed altogether.

Figs.~\ref{rcinf} and \ref{rcdm} show the final rotation curves for an infall
(I5) and dwarf merger (M1) simulation respectively.  The counterrotating gas
is colder than the primary stars.  The rotation curve of the counterrotating
gas is slightly higher than that of the primary, and its velocity dispersions
are systematically lower than those of the primary stars.  This is
qualitatively in agreement with the observed rotation curves, but our velocity
dispersion values for the primary disk stars do not agree well with the
observed values reported by JBH.  Our values are significantly lower, in some
cases by as much as a factor of 2.  The velocity dispersion values for the
counterrotating gas agree with the observations much better.  Since our gas
never gets converted into stars, we do not have values for the velocity
dispersions for counterrotating stars.

\subsection{Comparison with Previous Work}
The heating experienced by our primary disk after mergers with dwarfs that
have half its own mass (not counting the dark matter) is modest compared with
that due to a dissipationless merger between a large disk galaxy like the
Milky Way, and a dense dwarf galaxy with one-tenth the mass of the primary
disk (\cite{wmh96}).  The latter causes the disk to thicken by $\sim60$\% at
the solar circle, which is not surprising considering that the merger is
dissipationless, the primary is cold and thin, and a good portion of the
satellite's mass ($\sim45$\%) survives intact and sinks to the center of the
primary.  It is also a prograde merger which completes rapidly, in less than 3
Gyr.  This is due to the interaction of the satellite with the disk particles,
caused by the small initial distance of the satellite from the disk (6 disk
scale lengths).  There is not enough time for the primary to completely
tidally strip the satellite and thereby transfer orbital energy to the tidal
remnants.  The halo also does not have the opportunity to absorb some of this
energy.  A large initial orbital momentum merger (in which the satellite
starts out at several disk radii) of a 10\% mass satellite, by contrast, is
much less damaging to the primary (\cite{hc96}).  This is more in agreement
with our simulations, in which the infalling gas as well as the dwarf galaxy
are initially placed at a large average distance ($\sim20$ disk radii) from
the primary disk.  The relatively low density of the satellite in each of our
mergers (central density of primary is $\sim40$ times the central density of
the dwarf) also contributes to more efficient stripping of the satellite.

The higher initial angular momentum and lower density satellites are only part
of the story, however.  As mentioned in \S{4.2}, the nature of the primary in
NGC~4138 has a lot to do with the amount of heating experienced.  The
fragility of cold and thin galactic disks has been demonstrated time and again
in the context of minor mergers by numerical (TR; \cite{wmh96}; \cite{qhf93};
\cite{qg86}) and analytical (\cite{to92}; \cite{qg86}) studies, but the
response of early-type disk galaxies like NGC~4138 to minor mergers has not
received much attention.  The resistance of these disks to minor mergers
appears to be significantly higher.  The production of counterrotating disks
and bulges via retrograde minor mergers is much more likely for S0s and Sas
than for later type spirals.  The counterrotating bulge recently reported in
the Sb galaxy NGC~7331 (\cite{pgp96}) is most likely a result of very low
angular momentum gas infall rather than a minor merger.

The origin of the gas for continuous infall is an open question, and for this
reason the dwarf merger process is potentially a more frequent producer of
counterrotating disks.  The formation of dwarf galaxies in tidal tails of
interactions between larger galaxies has numerical (\cite{bh92b}; \cite{ekt93}; 
\cite{hm95}) as well as observational (\cite{hgv94}; \cite{hcz96}) support.  It
provides a possible mechanism for obtaining gas-rich dwarf galaxies which have
less dark matter than the norm.  The tidal features observed (\cite{hgv94}) in
NGC~7252 (the ``Atoms for Peace'' galaxy), for instance, contain clumps with
properties typical of dwarf irregular galaxies in terms of their high atomic
gas content, luminosity, velocity width and $M/L$ ratios ($\sim1.5-7.5$ for
$50\leq H_{\circ}\leq 75$).  These tidally formed dwarfs, however, are
substantially less massive than the dwarfs used in simulations M1-M3.

The S0 galaxy NGC 7332 (\cite{fif94}) shows extended counterstreaming [O~III]
gas in which the counterrotating gas has higher velocity than the prograde gas
and stars.  This is cited as support for recent accretion and agrees with the
behavior seen in our infall simulation with prograde gas (I6), where we
find that the counterrotating gas has higher angular momentum than the
corotating gas.  We do not, however, see any clear gas motion towards the
center as a result of the interaction of the counterstreaming components.  An
analytical study of two counterrotating gas disks lying one over the other
predicts that the counterrotating gas should drag the corotating gas inward
with equal masses of corotating and counterrotating gas accreting rapidly
(\cite{lc96}).  Although we see enhanced accretion for infall with prograde
gas (I6), the amount of prograde gas in our test is small and cannot be
expected to dramatically affect the accretion and inflow rate of the
counterrotating gas.  We do however see some evidence of the pile-up of gas
into rings at one or more radial distances predicted by \cite{ljh96} as a
consequence of a two-stream instability in counterrotating systems.  This is
evident in Fig.~\ref{rhgcmp}, especially for M2 and I6, where local peaks
and troughs dominate the radial gas profile.

The evolution of the tidal material seen in our simulations is in general
agreement with the behavior described by \cite{hi95} as a generic result of
mergers.  For the lowest orbital angular momentum case (I3), we see the tidal
material falling back early in shells, and as the orbital angular momentum
increases, so does the fallback time as the tidal material forms larger and
larger loops.  We also confirm that all the tidal material remains bound and
eventually will fall back into the primary galaxy.  Nested, counterrotating,
stable bars within primary bars have been observed in simulations of
retrograde gas-rich dwarf mergers where the primary is an early-type barred
galaxy (\cite{fr96}), but we do not see this happening in our simulations.
One reason for this is that there is not much counterrotating gas collecting
in the center of the primary disk.

\section{Conclusions}
Based on numerical simulations carried out using a fiducial spiral galaxy and
infall/merger model, TR predicted that counterrotating disks should be a rare
occurence because they require special conditions which must persist over
several Gyrs.  This was partly due to the fact that dwarf merger simulations
were not able to produce counterrotating disks with the required properties
within the required time frame.  Our experiments with NGC~4138 lead us to be
more optimistic about the chances of forming counterrotating disks with dwarf
mergers.  The combined effects of a massive, compact primary and a smaller
dynamical time (due to the smaller size of the primary) gives a dwarf merger a
much better chance at success.  For a dwarf merger model based on the
observationally determined parameters of the galaxy, we have found a
configuration which produces a counterrotating disk closely matching the one
observed, although it does require the dwarf galaxy to be predominantly
gaseous with little to moderate amounts of dark matter.

The results of minor mergers involving large primary galaxies with cold, thin
disks cannot be extended to all types of spirals.  Early type spirals, with
hot, massive and compact disks of the type that NGC~4138 has, appear to be
more resistant and less fragile.  The initial orbital angular momentum of the
satellite plays a role in determining the damage done to the primary disk as
well in numerical simulations, and it must be taken into account when
interpreting the results.

Continuous infall also succeeds in producing a counterrotating disk which
matches the observed one in terms of mass and size.  The time frame for infall
is somewhat longer than for dwarf mergers.  Whereas a dwarf merger yields a
counterotating disk in $3-4$ Gyr, the infall process accomplishes this in
$\sim5$ Gyr.

A common characteristic of the counterrotating gas disk resulting from both
infall and mergers is that it does not have an exponential profile like the
primary disk.  This is not a problem for NGC~4138, since there is no evidence
of its counterrotating disk having an exponential profile, but counterrotating
disks in some other galaxies do have exponential profiles (e.g., NGC~4550,
NGC~3593).

The nuclear star formation ring may owe its origin to an interaction
between prograde and retrograde gas or to a pile up of gas at a resonance.
The counterstreaming gas hypothesis is viable for both infall and mergers.
The inferred star formation histories of the primary and counterrotating disks
do not pose serious problems for this hypothesis currently, but knowledge of
the colors and ages of the counterrotating stars is necessary to resolve this
issue.

As pointed out by JBH, we have only tested a restricted set of infall and
merger possibilities so far.  This is partly because these processes are not
well constrained observationally (especially infall), but also because of our
CPU resource limitations.  There is a need to explore alternative infall and
merger models before a reliable prediction can be made regarding the overall
ease of obtaining massive CR disks in all types of spiral galaxies.

\acknowledgments
A.R.T. thanks Joshua Barnes for providing the C version of the basic tree
code, and Richard Pogge and Glenn Tiede for useful discussions.  We thank
Richard Pogge for comments on the manuscript.  Support for this work was
provided by a NYI award to B.S.R. (NSF grant AST-9357396), NASA grant NAG
5-2864, and an Ohio Supercomputer Center Research Grant (PAS825) for computer
time on the Cray Y-MP.  We wish to thank the O.S.C. staff for their
assistance.  Finally, we thank the referee, Chris Mihos, for providing a
thorough and insightful referee report.

\clearpage
\begin{figure}
\caption{The initial rotation curves for the primary disk and halo in our
numerical model compared with the corresponding rotation curves derived from
observations of NGC~4138 (JBH).}
\label{rcinit}
\end{figure}

\begin{figure*}
\caption{Gas infall with large initial angular momentum for the gas (I1).  The
top half shows the ``top'' view and the bottom half shows the ``side'' view.
The primary disk is shown only in the first panel for each view for clarity.
The galactic halo is not shown.  The mass of the gas is 50\% of the mass of
the disk. The square panels are 80 kpc on each side, except for the last
panel, which is 40 kpc on each side.  Time is in Gyr.}
\label{ci1}
\end{figure*}

\begin{figure*}
\caption{Top (upper half) and side (lower half) view of gas infall with just
just enough initial angular momentum to still yield a counterrotating disk
(I2).  Panel widths are 80 kpc except for the last panel in each view, which
is 40 kpc wide.}
\label{ci2}
\end{figure*}

\begin{figure*}
\caption{Top (upper half) and side (lower half) view of gas infall with very
low initial angular momentum (I3).  Panel widths are 80 kpc except for the
last panel in each view, which is 40 kpc wide.}
\label{ci3}
\end{figure*}

\begin{figure*}
\caption{Velocity fields of the disk (bottom) and counterrotating gas (top)
compared for I3 (left) vs. I1 (right).  Only a fraction of the particles
are sampled, and sampling is proportional to radius.}
\label{vfcmp}
\end{figure*}

\begin{figure}
\caption{The mean specific angular momentum of the disk and counterrotating   
gas compared for I3 (open squares) and I1 (filled squares).  In our sign 
convention, $J_z>0$ for prograde orbits and $J_z<0$ for retrograde orbits.}
\label{amcmp}
\end{figure}

\begin{figure*}
\caption{Top (far left) and side (left) views of the gas,  and top (right) and
side (far right) views of the primary disk for gas infall with an orbital
inclination of $\sim30^{\circ}$ (I4).  Panel widths are in kpc.}
\label{ci4}
\end{figure*}

\begin{figure*}
\caption{Top (left) and side (center) views of the gas, and side views of the
primary disk (right), for continuous gas infall with a smaller search radius
for sticky particle neighbors (I5).  The primary disk is shown for reference
also in the first panel of each of the gas views.  For the gas, the square
panels are 80 kpc wide, with the last one in each view being 20 kpc wide.  For
the primary disk, all panels are 20 kpc wide.}
\label{ci5}
\end{figure*}

\begin{figure*}
\caption{Top (upper half) and side (lower half) view of continuous gas infall
with a prograde ring of gas in the primary disk initially at $\sim1.2$ kpc
from the center (I6).  The prograde gas is also shown.  The last two panels
show magnified views of the counterrotating disk.  The square panels are 80
kpc wide, except for the last one, which is 20 kpc wide.}
\label{ci6}
\end{figure*}

\begin{figure*}
\caption{Top (upper half) and side (lower half) views of a gas-rich dwarf
merger with very little dark matter in the dwarf (M1).  The primary disk is
shown as light grey, and both the stars and the gas in the dwarf are shown as
dark particles.  The dark halos are not shown.  The square panels are 80 kpc
wide, except for the last panel, which is 40 kpc wide.}
\label{dm1}
\end{figure*}

\begin{figure*}
\caption{Top (upper half) and side (lower half) views of a gas-rich dwarf
merger (M2) with 4 times the amount of dark matter as in M1.}
\label{dm2}
\end{figure*}

\begin{figure*}
\caption{A comparison of the two dwarf mergers M1 and M2, showing three
species: the primary disk (left), the gas in the dwarf (center) and the stars
in the dwarf (right).  The angular momentum comparison plots are in the middle
(open squares for M1, filled squares for M2), with the velocity fields and
the side views for M1 above the A.M. plots, and the velocity fields and side
views for M2 below them.}
\label{dmcmp}
\end{figure*}

\begin{figure}
\caption{A comparison of the thickness of the disk after continuous infall 
and dwarf mergers.  The solid line shows the initial disk thickness at $t=0$.
The dotted line is the disk thickness at $t=4.5$ Gyr for an isolated disk
evolved without any infall or merger.  Thicknesses for $R\gtrsim4$ kpc are not
reliable due to small numbers of particles at larger disk radii.}
\label{zthcmp}
\end{figure}

\begin{figure}
\caption{A comparison of the radial mass distribution of the counterrotating
disk for infall and dwarf mergers.  The $y$-axis shows the number density $N$
of gas particles.}
\label{rhgcmp}
\end{figure}

\begin{figure}
\caption{A comparison of the mean specific angular momentum of the
counterrotating gas (open squares) with that of the corotating (prograde) gas 
for I6.}
\label{ampro}
\end{figure}

\begin{figure}
\caption{The final rotation curves and velocity dispersion plots for the
primary disk (filled points) and the counterrotating gas (open points) after
gas infall (I5).  The initial disk rotation curve is drawn in for
comparison.  The velocity dispersion values are somewhat smaller than the
observed values.  The disk inclination is taken into account in calculating
the rotation speed.}
\label{rcinf}
\end{figure}

\begin{figure}
\caption{The final rotation curves and velocity dispersion plots for the
primary disk (filled points) and the counterrotating gas (open points) for the
dwarf merger M1.  The initial disk rotation curve is drawn in for comparison.
Both the rotation curves and the velocity dispersion values are higher than
those for the infall simulation, although part of the difference may be due to
the error in the inclination correction.  The dispersion values are closer to
the observed values in this case.}
\label{rcdm}
\end{figure}

\clearpage
\begin{table}
\caption{PARAMETERS FOR NGC~4138 MODEL.}
\begin{tabular}{cccccccc}
\\ \tableline\tableline
\multicolumn{4}{c}{Disk} & & \multicolumn{3}{c}{Halo} \\
\cline{1-4} \cline{6-8} \\
Mass & Radius & Scale Len. & Scale Ht. & & Mass & Radius & Core
Rad. \\ $M_d$ & $R_D$ & $R_d$ & $h_d$ & & $M_h$ & $R_h$ & $R_c$ \\ \tableline
$2.0\times10^{10}$ & 6.0 & 1.25 & 0.2 & & $1.5\times10^{11}$ & 20 & 5 \\
\tableline
\end{tabular}
\label{dhpars}
\tablecomments{Masses are in $M_{\sun}$, lengths in kpc.}
\end{table}

\begin{table}
\caption{PARAMETERS FOR GAS INFALL SIMULATIONS.}
\begin{tabular}{lcccccc}
\\ \tableline\tableline
Model & $M_g\tablenotemark{a}$ & $V_0\tablenotemark{b}$ &
$X_0\tablenotemark{c}$ & $J_z\tablenotemark{d}$ & $r_{\rm
nn}\tablenotemark{e}$ & $Z_0\tablenotemark{f}$
\\ \tableline 
I1 & $1.0\times10^{10}$ & 0.3 & 25 & 0.45 & 3.0 & 0\\
I2 & $1.0\times10^{10}$ & 0.3 & 10 & 0.18 & 3.0 & 0 \\
I3 & $1.0\times10^{10}$ & 0.2 & 10 & 0.12 & 3.0 & 0 \\
I4 & $1.0\times10^{10}$ & 0.3 & 20 & 0.36 & 3.0 & 15 \\
I5 & $1.0\times10^{10}$ & 0.3 & 25 & 0.45 & 1.0 & 0 \\
I6 & $1.1\times10^{10}$\tablenotemark{\dag} & 0.3 & 25 & 0.45 & 1.0 & 0 
\\ \tableline
\end{tabular}
\label{infpars}
\tablenotetext{\rm a}{total gas mass, in $M_{\sun}$}
\tablenotetext{\rm b}{the initial velocity of the gas in the negative $y$
direction, specified in units of the centripetal velocity 
$\sqrt{G(M_h+M_d)/X_0}$, where $M_h+M_d$ is the total mass of the primary
(halo+disk)}
\tablenotetext{\rm c}{the initial distance of the gas column from the center of
the primary disk, in kpc}
\tablenotetext{\rm d}{the magnitude of the angular momentum (quantified crudely
by taking the product $m_g\times V_0\times R_0$, where $m_g$ is the total mass
of the infalling gas in our simulation units)}
\tablenotetext{\rm e}{the search radius for nearest neighbors used by the sticky
particle code, in kpc}
\tablenotetext{\rm f}{the initial height of the gas column above the $x-y$ plane,
in kpc}
\tablenotetext{\rm \dag}{I6 has additional prograde gas in the primary disk with
one-tenth the mass of the infalling gas.}
\end{table}

\begin{table}
\caption{DWARF GALAXY PARAMETERS FOR GAS-RICH DWARF MERGERS.}
\begin{tabular}{lccccccc}
\\ \tableline\tableline
Model & $M_g$\tablenotemark{a} & $M_*$\tablenotemark{b} &
$M_D$\tablenotemark{c} & $M_g/M_T$\tablenotemark{d} & $R_*$\tablenotemark{e}
& $R_g$\tablenotemark{f} & $R_h$\tablenotemark{g} \\ \tableline  
M1 & $8\times10^9$ & $2\times10^9$ & $5\times10^9$ & 0.53 & 4.0 & 4.0 & 4.0 \\
M2 & $8\times10^9$ & $2\times10^9$ & $2\times10^{10}$ & 0.27 & 4.0 & 4.0 & 8.0 \\
M3\tablenotemark{h} & $8\times10^9$ & $2\times10^9$ & $2\times10^{10}$ & 0.27 & 4.0 & 4.0 & 8.0 \\ \tableline
\end{tabular}
\label{dmpars}
\tablenotetext{\rm a}{mass of gas, in $M_{\sun}$}
\tablenotetext{\rm b}{mass of stars, in $M_{\sun}$}
\tablenotetext{\rm c}{mass of dark matter, in $M_{\sun}$}
\tablenotetext{\rm d}{ratio of gas mass to total mass}
\tablenotetext{\rm e}{radius of stellar sphere, in kpc}
\tablenotetext{\rm f}{radius of gas sphere, in kpc}
\tablenotetext{\rm g}{radius of halo, in kpc}
\tablenotetext{\rm h}{M3 uses twice the number of particles as M2}
\end{table}


\begin{thebibliography}{}
\bibitem[Balcells \& Quinn 1990]{bq90} Balcells, M., \& Quinn, P.J. 1990, \apj, 361, 381. 
\bibitem[Barnes 1992]{ba92} Barnes, J.E. 1992, \apj, 393, 484. 
\bibitem[Barnes \& Hernquist 1991]{bh91} Barnes, J.E., \& Hernquist, L.E. 1991, \apj, 370, L65. 
\bibitem[Barnes \& Hernquist 1992a]{bh92a} Barnes, J.E., \& Hernquist, L.E. 1992a, \araa, 30, 705.  
\bibitem[Barnes \& Hernquist 1992b]{bh92b} Barnes, J.E., \& Hernquist, L.E. 1992b, Nature, 360, 715. 
\bibitem[Barnes \& Hut 1986]{bh86} Barnes, J.E., \& Hut, P. 1986, Nature, 324, 446. 
\bibitem[Bertola et al. 1996]{bc96} Bertola, F., Cinzano, P., Corsini, E.M.,
\& Pizzella, A. 1996, \apj, 458, L67. 
\bibitem[Binney \& Tremaine 1987]{bt87} Binney, J., \& Tremaine, S. 1987, Galactic Dynamics (Princeton University Press), \S7.1 and eq. 7-27.
\bibitem[Brahic 1977]{br77} Brahic, A. 1977, \aap, 54, 895. 
\bibitem[Braun, Walterbos \& Kennicutt 1992]{bwk92} Braun, R., Walterbos, R.A.M., \& Kennicutt, R.C. Jr 1992, Nature, 360, 442.
\bibitem[Ciri, Bettoni \& Galletta 1995]{cb95} Ciri, R., Bettoni, D., \&
Galletta, G. 1995, Nature, 375, 661.
\bibitem[de Vaucouleurs et al. 1991]{dv91} de Vaucouleurs, G., de Vaucouleurs,
A., Corwin, H.G., Buta, R.J., Paturel, G., \& Fouque, P. 1991, Third Reference
Catalogue of Bright Galaxies (Springer, New York).
Galletta, G. 1995, Nature, 375, 661.
\bibitem[Elmegreen, Kaufmann \& Thomasson 1993]{ekt93} Elmegreen, B.,
Kaufmann, M. \& Thomasson, M. 1993, \apj, 412, 90. 
\bibitem[Fisher, Illingworth \& Franx 1994]{fif94} Fisher, D.,  Illingworth,
G. \& Franx, M. 1994, \aj, 107, 160. 
\bibitem[Friedli 1996]{fr96} Friedli, D. 1996, \aap, in press.
\bibitem[Galletta 1987]{ga87} Galletta, G. 1987, \apj, 318, 531.
\bibitem[Hernquist \& Barnes 1991]{hb91} Hernquist, L.E., \& Barnes, J.E. 1991, Nature, 354, 210. 
\bibitem[Hibbard (1995)]{hi95} Hibbard, J.E. 1995, Ph.D. Thesis, Columbia
University, Chs. IV and VI.
\bibitem[Hibbard et al. 1994]{hgv94} Hibbard, J.E., Guhathakurta, P., van
Gorkom, J.H., \& Schweizer, F. 1994, \aj, 107, 67.
\bibitem[Hibbard \& Mihos 1995]{hm95} Hibbard, J.E., \& Mihos, J.C. 1995, \aj,
110, 140.
\bibitem[Huang \& Carlberg 1996]{hc96} Huang, S. \& Carlberg, R.G. 1996, \apj, submitted.
\bibitem[Hunsberger, Charlton \& Zaritsky 1996]{hcz96} Hunsberger, S.D.,
Charlton, J.C., \& Zaritsky, D. 1996, \apj, submitted.
\bibitem[Jore, Broeils \& Haynes 1996]{jbh96} Jore, K.P., Broeils, A.H., \&
Haynes, M.P. 1996, \apj, submitted. [JBH]
\bibitem[Kuijken, Fisher \& Merrifield 1996]{kfm96} Kuijken, K., Fisher, D., \&
Merrifield, M.R. 1996, \mnras, in press.
\bibitem[Lovelace \& Chou 1996]{lc96} Lovelace, R.V.E., \& Chou, T. 1996, \apj, submitted.
\bibitem[Lovelace, Jore \& Haynes (1996)]{ljh96} Lovelace, R.V.E., Jore, K.P., \&
Haynes, M.P. 1996, \apj, submitted.
\bibitem[Merrifield \& Kuijken 1994]{mk94} Merrifield, M.R., \& Kuijken, K. 1994, \apj, 432, 575.
\bibitem[Negroponte \& White 1983]{nw83} Negroponte, J., \& White, S.D.M. 1983, \mnras, 208, 1009. 
\bibitem[Pogge \& Eskridge 1987]{pe87} Pogge, R.W. \& Eskridge, P.B. 1987, \aj, 93, 291.
\bibitem[Pogge \& Eskridge 1993]{pe93} Pogge, R.W. \& Eskridge, P.B. 1993, \aj, 106, 1405.
\bibitem[Prada et al. 1996]{pgp96} Prada, F., Gutierrez, C.M., Peletier, R.F.,
\& McKeith C.D. 1996, \mnras, submitted.
\bibitem[Quinn \& Goodman 1986]{qg86} Quinn, P.J. \& Goodman, J. 1986, \apj, 309, 472.
\bibitem[Quinn, Hernquist \& Fullagar 1993]{qhf93} Quinn, P.J., Hernquist, L.E., \& Fullagar, D.P. 1993, \apj, 403, 74.
\bibitem[Roberts \& Haynes 1994]{rh94} Roberts, M.S. \& Haynes, M.P. 1994, \araa, 32, 115. 
\bibitem[Rubin, Graham \& Kenney 1992]{rgk92} Rubin, V.C., Graham, J.A., \& Kenney, J.D.P. 1992, \apj, 394, L9.
\bibitem[Sage \& Galletta 1994]{sg94} Sage, L.J. \& Galletta, G. 1994, \aj, 108, 1633. 
\bibitem[Sandage \& Tammann 1981]{st81} Sandage, A.R., and Tammann, G.A. 1981,
A Revised Shapley-Ames Catalog of Bright Galaxies, (Washington, Carnegie
Institution of Washington).
\bibitem[Schwarz 1981]{sw81} Schwarz, M.P. 1981, \apj, 247, 77. 
\bibitem[Toth \& Ostriker 1992]{to92} Toth, G. \& Ostriker, J.P. 1992, \apj, 389, 5. 
\bibitem[Thakar \& Ryden 1996]{tr96} Thakar, A.R., \& Ryden, B.S. 1996, \apj,
461, 55. [TR]
\bibitem[Thakar 1996]{th96} Thakar, A.R. 1996, \baas, 28, 959.
\bibitem[Walker, Mihos \& Hernquist 1996]{wmh96} Walker, I.R., Mihos, J.C., \&
Hernquist, L.E. 1996, \apj, 460, 121. 
\end{thebibliography}
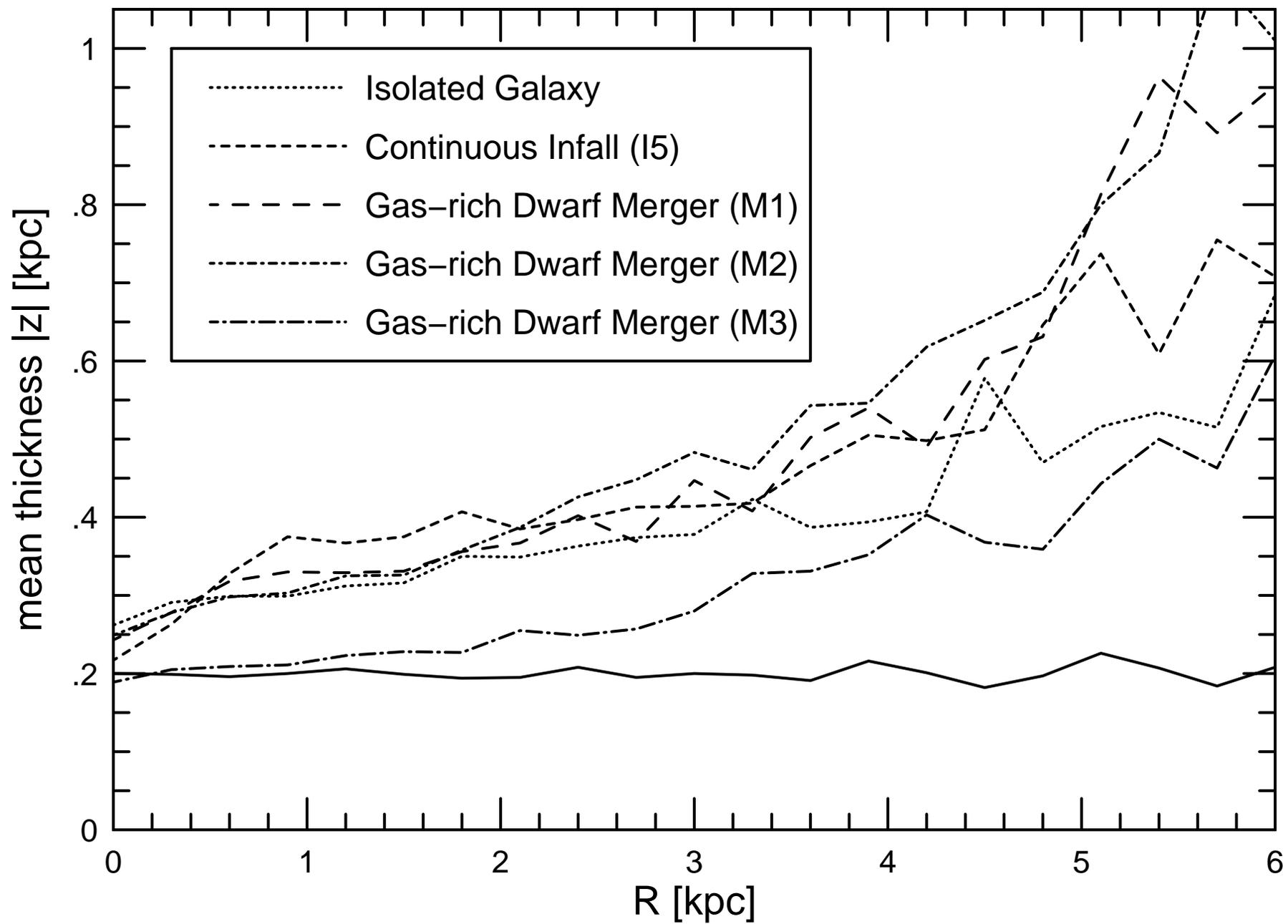
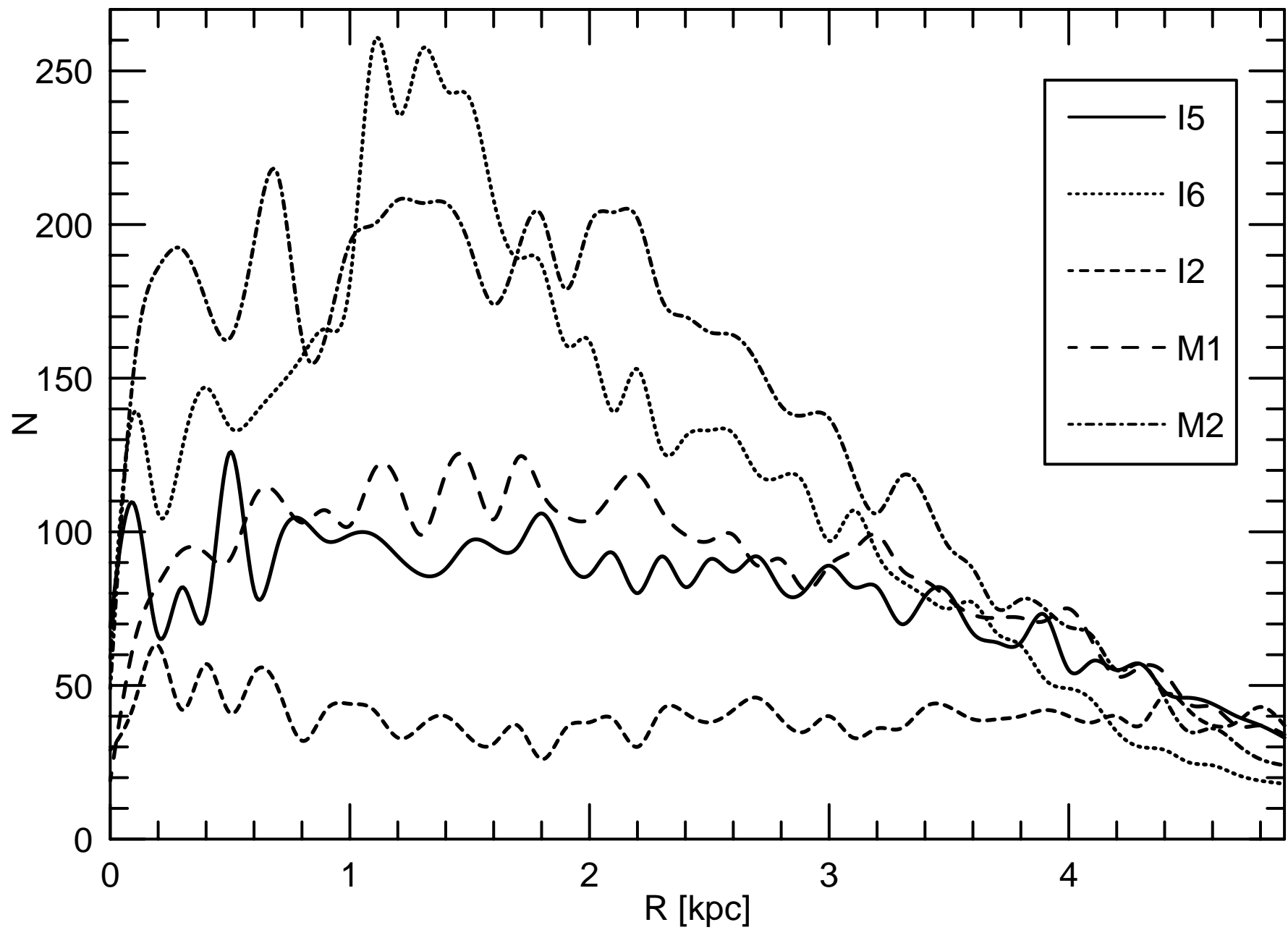
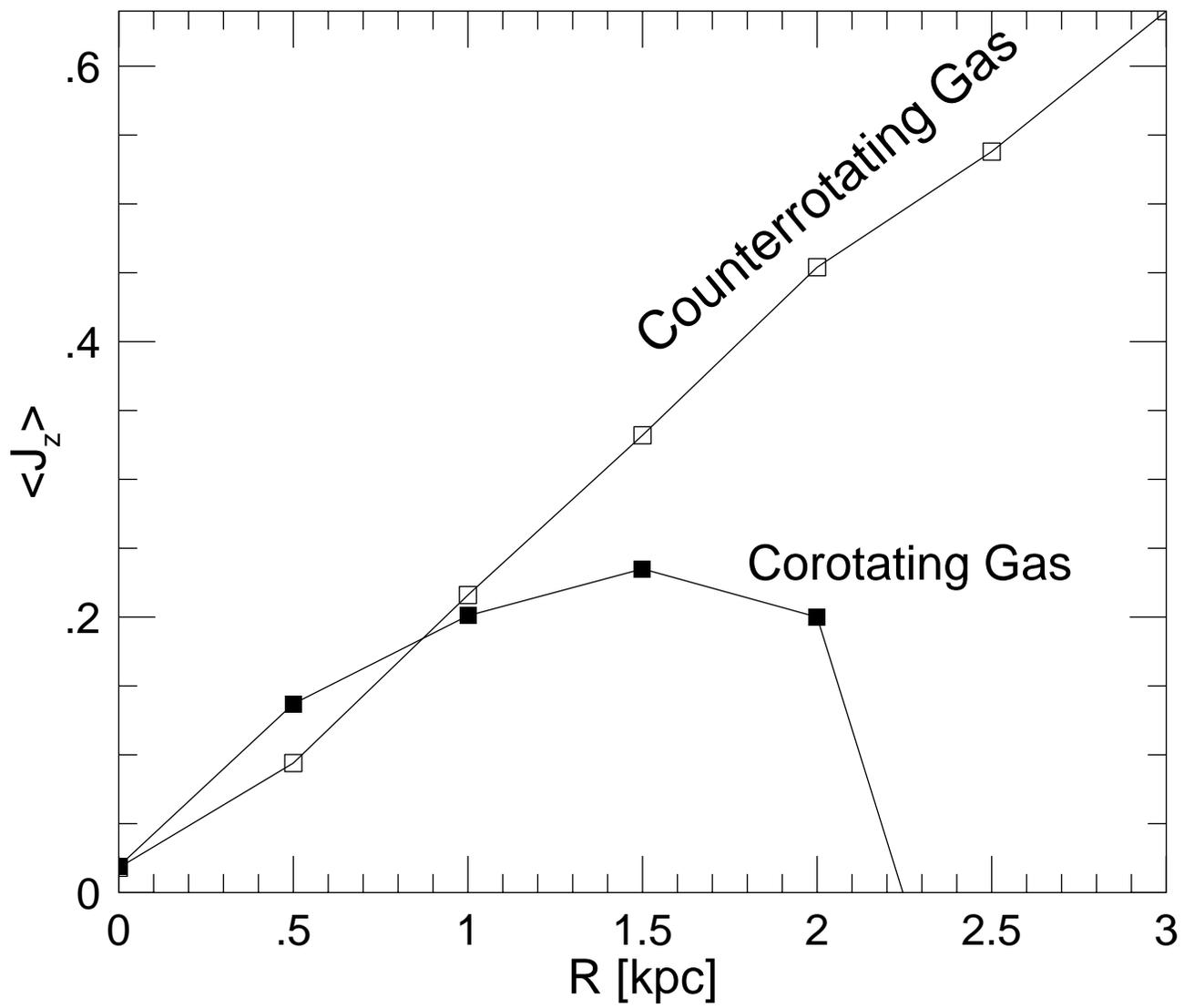
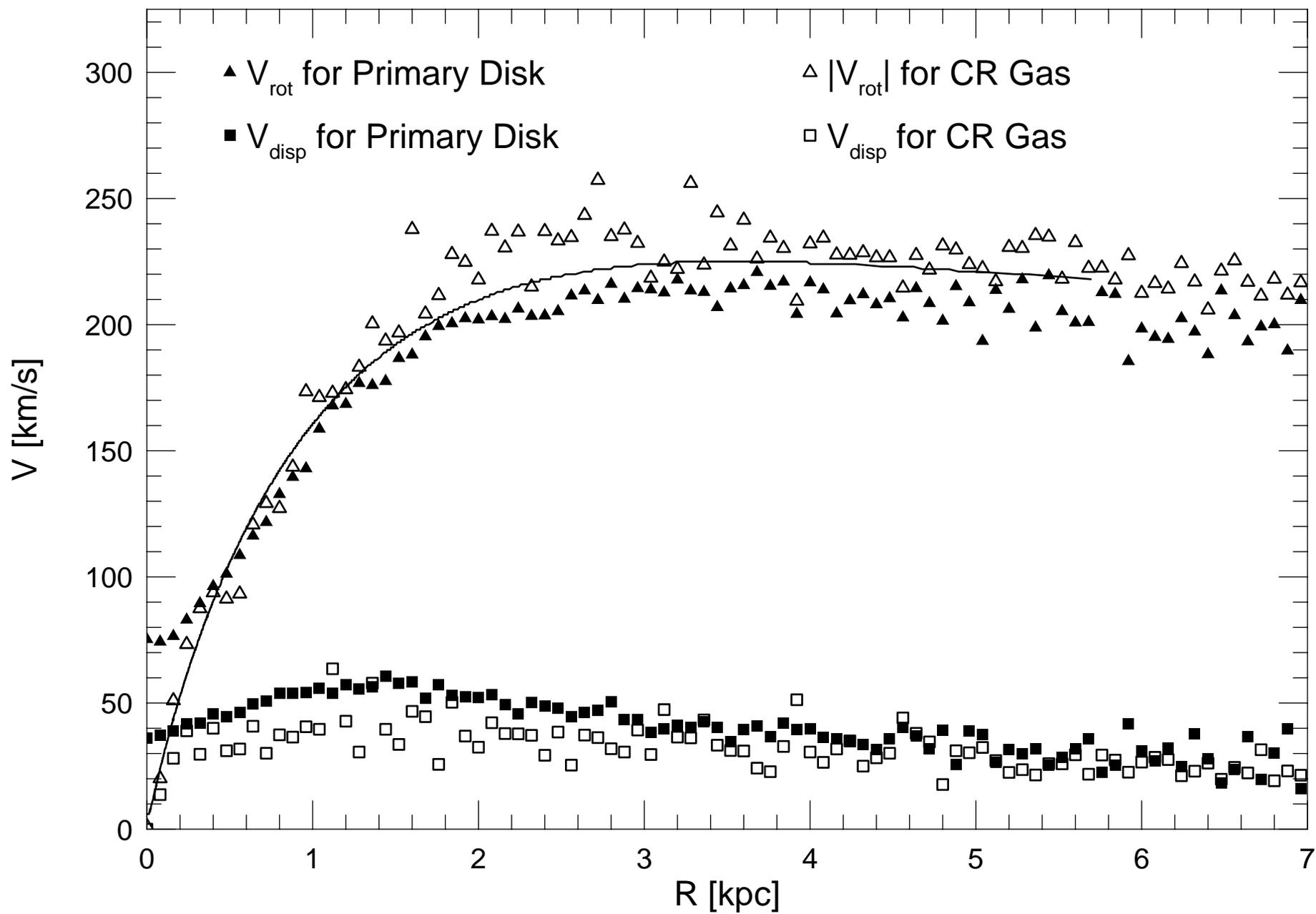
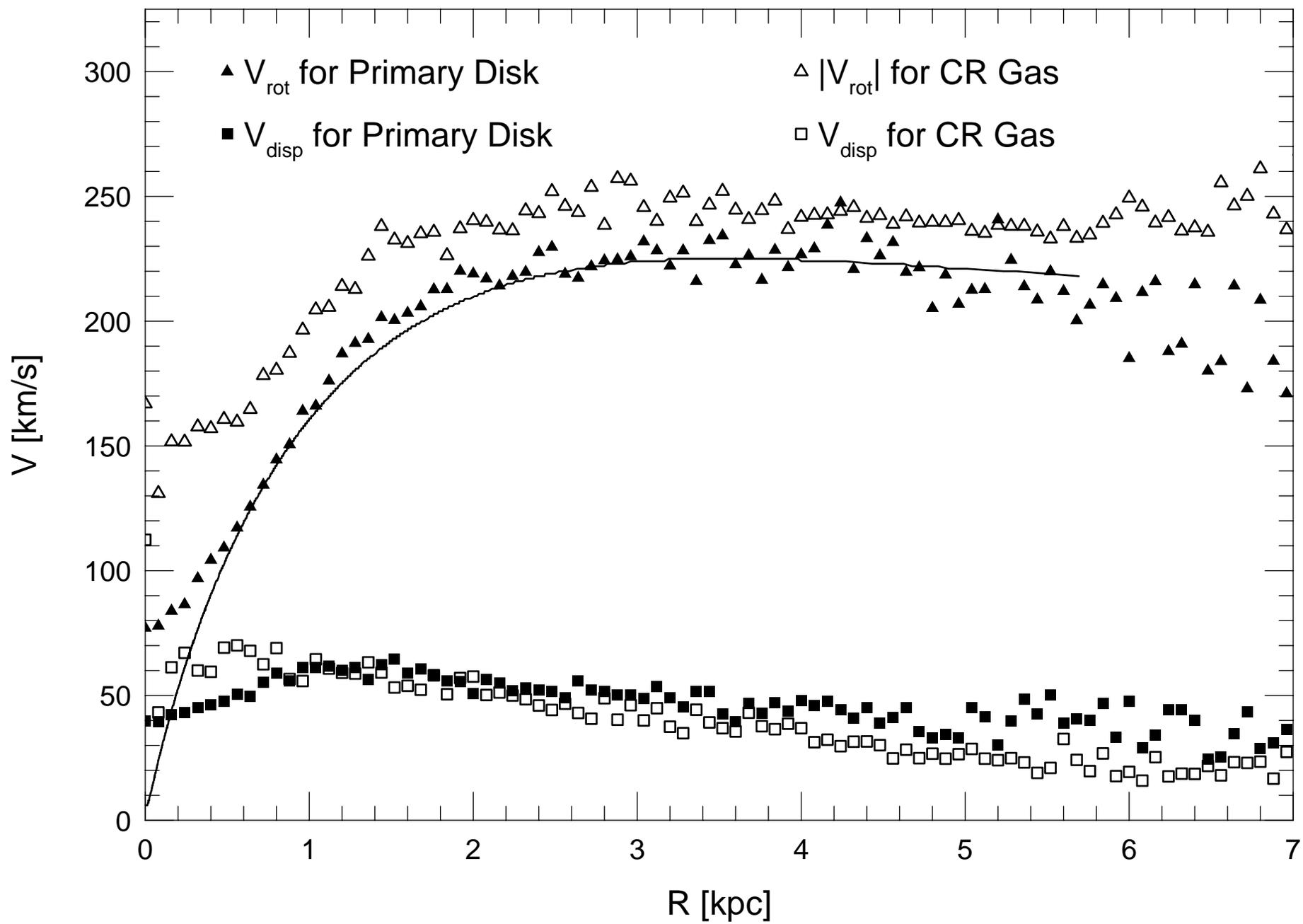
\end{document}